\documentclass[FBSedit,FBSmath,ecsub]{FBSsuppl}
\usepackage{amsfonts}
\usepackage{amssymb}
\usepackage{epsfig}
\newcommand{\ccdot}{ \hspace{-0.8mm} \cdot \hspace{-0.8mm}}

\title{Nucleon-Nucleon Interaction Models and Non-Locality}
\author{B. Desplanques\instnr{1}\thanks{\textit{E-mail address:} 
desplanq@isn.in2p3.fr}, A. Amghar\instnr{2}}
\instlist{Institut des Sciences Nucl\'eaires (UMR CNRS/IN2P3--UJF),  
  F-38026 Grenoble Cedex, France 
\and
Facult\'e  des Sciences, Universit\'e de Boumerdes, 35000 Boumerdes, 
Algeria }

\begin{document}

\maketitle
\begin{abstract}
The effect of non-locality in the NN interaction models is examined. It is shown 
that this feature can explain differences in predictions made from models 
evidencing a difference with this respect. This is done for both static and 
dynamical observables, taking into account that a non-local term can be 
transformed away by performing a unitary transformation. Some results for the 
deuteron form factors, the $A(Q^2)$ structure function and the $T_{20}(Q^2)$ 
tensor polarization are given as an example. A few cases where discrepancies 
cannot be explained are also considered. They point to differences in the models 
as for the deuteron asymptotic normalizations, $A_S$ and $A_D$, which are not 
affected by the present analysis.
\end{abstract}

\section{Introduction}
Apart from a recent work by Doleschall and Borb\'ely\cite{DOLE}, non-locality in 
the NN interaction has not been the object of dedicated studies for a long time. 
It is not absent however in models where it appears most often as a by-product 
of some prejudice in their construction. In the Paris model\cite{PARI}, for 
instance, it was realized that an energy dependence  could help in fitting NN 
scattering data. The transformation of this energy dependence into a $p^2/M$ 
dependence provides a non-local component. Later on, the Bonn group produced a 
model, field-theory motivated, taking into account the coupling of the $NN$ 
channel to $NN\pi$, N$\Delta$, $\Delta\,\Delta,\cdots$ channels\cite{BONN}. It 
contains both a spatial and a time-non-locality. Moreover, the improvement 
consisting in introducing the Dirac spinors to describe $1/2$-spin particles 
also provides non-locality. One could add other examples that have not been 
concretized in a high accuracy model. Taking into account the substructure of 
nucleons and mesons in terms of quarks most often leads to a 
non-locality\cite{KUKU},  which is better expressed in configuration space than 
in momentum one, contrarily to the above sources of non-locality.  

A double question may be raised about this non-locality. Does it help in 
explaining NN scattering data and how this could be evidenced? On the other 
hand, taking into account that a non-locality can be transformed away by a 
unitary transformation (wave by wave), one can wonder whether the different 
models on the market are independent of each other?

Concerning the first question, some answer is obtained by examining models such 
as the versions Nij1 (non-local) and Nij2 (local) of the Nijmegen 
group\cite{STOK}. They equally fit the scattering phase shifts ($\chi^2 \;{\rm 
per \;datum}= 1.03$), but in the first case, this is achieved with 41 parameters 
while 47 are required in the other one. The slightly smaller number in the 
former case perhaps provides indication that the introduction of some 
non-locality is beneficial. 

The second question is the main object of the present paper. The plan will be as 
follows. In the second section, we present the different 
non-local terms which we are interested in. How they can be removed by a unitary 
transformation at the first order is given. The third section is devoted to a 
few selected results concerning the deuteron: static properties, form factors, 
structure function ($A(Q^2)$) and the tensor polarization  ($T_{20}(Q^2)$). It 
involves a comparison of these quantities obtained with different models when 
the effect of the non-locality is taken into account. Section four contains the 
conclusion and a discussion. Due to a lack of space, we concentrate here on the 
essential points. Details and extended results could be found in refs. 
\cite{DESP1,AMGH0,AMGH1}.

\section{Transforming Away Non-Local Terms}
The interaction of interest here may be written:
\begin{equation}
V=V_S+V_T+ \{\frac{\bf p^2}{M},\tilde{W}_S\}
+ \{\frac{\bf p^2}{M}, \tilde{W}_T\}
+[\frac{\bf p^2}{M}, i\,U]. \label{def}
\end{equation}
where the non-local terms take the form of an anticommutator or a commutator. 
Another term of same order could be considered but those retained here are the 
only ones appearing in the pion-exchange contribution when this one is expanded 
up to order $1/M^4$. Due to its long range, it a priori provides larger 
contributions. Moreover, they are theoretically well identified while shorter 
range contributions are likely to have some effective character. The last term 
in the above equation has been studied at length in refs. \cite{DESP1,AMGH0, 
FORE}. Though it has a different origin, it has a strong similarity with a term 
arising from the difference of pseudo-scalar and pseudo-vector $\pi NN$ 
couplings, considered in an earlier work by Friar\cite{FRIA}. As for the  
anticommutator terms, only rough estimates were made in the past. They are 
considered more completely here. 

In principle, if two models are unitary equivalent, the corresponding 
Hamiltonians, $H$ and $H'$, should 
fulfill the following relation:
\begin{equation}
H=\frac{p^2}{M}+V+V_{NL}= e^{-S}\;H'\;e^S=
e^{-S}\; (\frac{p^2}{M}+V')\;e^S.
\label{unit}
\end{equation}
At the first order in the interaction, the quantity, $S$, appearing in the 
unitary transformation, ${\rm exp}(S)$, can be determined by requiring:
\begin{equation}
V_{NL}+[S,\frac{{\bf p^2}}{M}]= \Delta\,V^0 \; , 
\label{req}
\end{equation}
where $\Delta\,V^0$ has to be local. This equation is fulfilled as  
follows:
\begin{eqnarray}
S&=&iU 
+\frac{i}{2} \Bigg(\vec{p} \ccdot \vec{r} \;\;\Big( V_0(r) +S_{12}(\hat{r}) \; 
V_1(r) \Big) 
 \nonumber \\    & & 
+\Big(\vec{\sigma_1} \ccdot \vec{p} \; \vec{\sigma_2} \ccdot \vec{r} +
     \vec{\sigma_2} \ccdot \vec{p} \; \vec{\sigma_1} \ccdot \vec{r} - 
\frac{2}{3} \;
     \vec{\sigma_1} \ccdot \vec{\sigma_2} \;\vec{p} \ccdot \vec{r} \Big) \; 
V_2(r)
+  h.c. \Bigg),
\end{eqnarray} 
\vspace{-8mm}
with
\begin{eqnarray}
V_0(r)&=&-\frac{1}{r}\int_r^\infty W_S(r')dr', \nonumber \\ 
 V_2(r)&=&-\frac{1}{r}\int_r^\infty dr'\int_{r'}^\infty 
\frac{W_T(r'')}{r''}dr'',
\nonumber \\
V_1(r)&=&-V_2(r)-\int_r^\infty \frac{W_T(r')}{r'}dr'.
\label{sol}
\end{eqnarray}
It is noticed that the difference of $V$ and $V'$ in Eq. (\ref{unit}) 
involves two-body but also 
many-body terms. Similarly, if a one-body current is introduced in one 
representation, the other one contains two, ... body currents to ensure the 
unitary equivalence. The role of the two-body part is examined in the 
following section. Applications involve the interaction models:
Nij2\cite{STOK}, Argonne V18\cite{WIRI}, Reid93\cite{STOK} which are local ones,
Nij1\cite{STOK}, Nij93\cite{STOK}, Paris\cite{PARI} which have a linearly 
$p^2/M$ dependence, and the Bonn-QB, Bonn-CD  ones\cite{BONN}.

\section{Results for Static and Dynamical Quantities}
Concerning static properties, a quantity of interest is the deuteron D-state 
probability, often referred to characterize different models. The difference 
between the Paris and Bonn-QB model is 0.78\%. As the two models have quite 
close values for the mixing parameters, $\epsilon_1$, their comparison is 
largely free of bias with this respect. Taking into account the effect of the 
non-local term, $V_{NL}$, explains 0.74\% (0.60\% and 0.14\% for the commutator 
and anti-commutator parts respectively). Notice that the change in the deuteron 
D-state probability just reflects the fact that this quantity is not an 
observable one. Similar results were reached by von Geramb et 
al.\cite{GERA}, using the inverse scattering problem methods. Another 
interesting quantity is the ratio $Q_D/(A_S\;A_D)$, which, contrary to the above 
one, is observable. The difference is 0.23 fm$^3$ while $V_{NL}$ explains 0.20 
fm$^3$. It is also instructive to look at  quantities that only depend on the 
scalar anticommutator part of $V_{NL}$. In this order, we compare the squared 
charge radius 
for the models Nij1 and Nij2. The models differ by an amount of 0.004 fm$^2$ 
while the effect of the non-locality is estimated to be around 0.001 fm$^2$. The 
apparent failure to explain the discrepancy in this case actually points to the 
difference of the models for the asymptotic normalization $A_S$, which as is 
well known, governs the size of the charge radius. This factor is not affected 
by the present analysis of non-local effects.

\begin{figure}[htb]
\begin{center}
\mbox{ \epsfig{ file=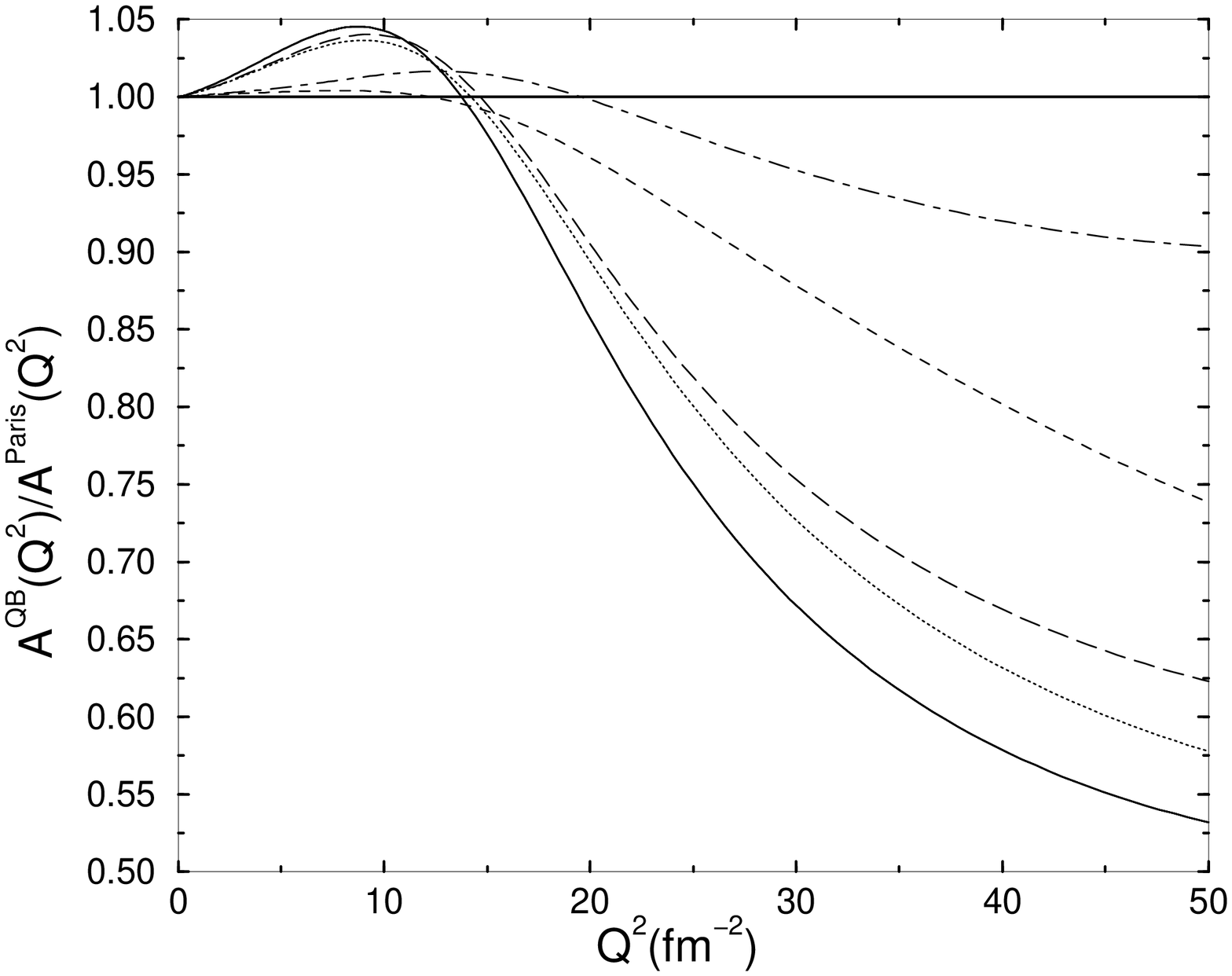, angle=270, width=5.5cm} 
\hspace*{0.5cm} \epsfig{ file=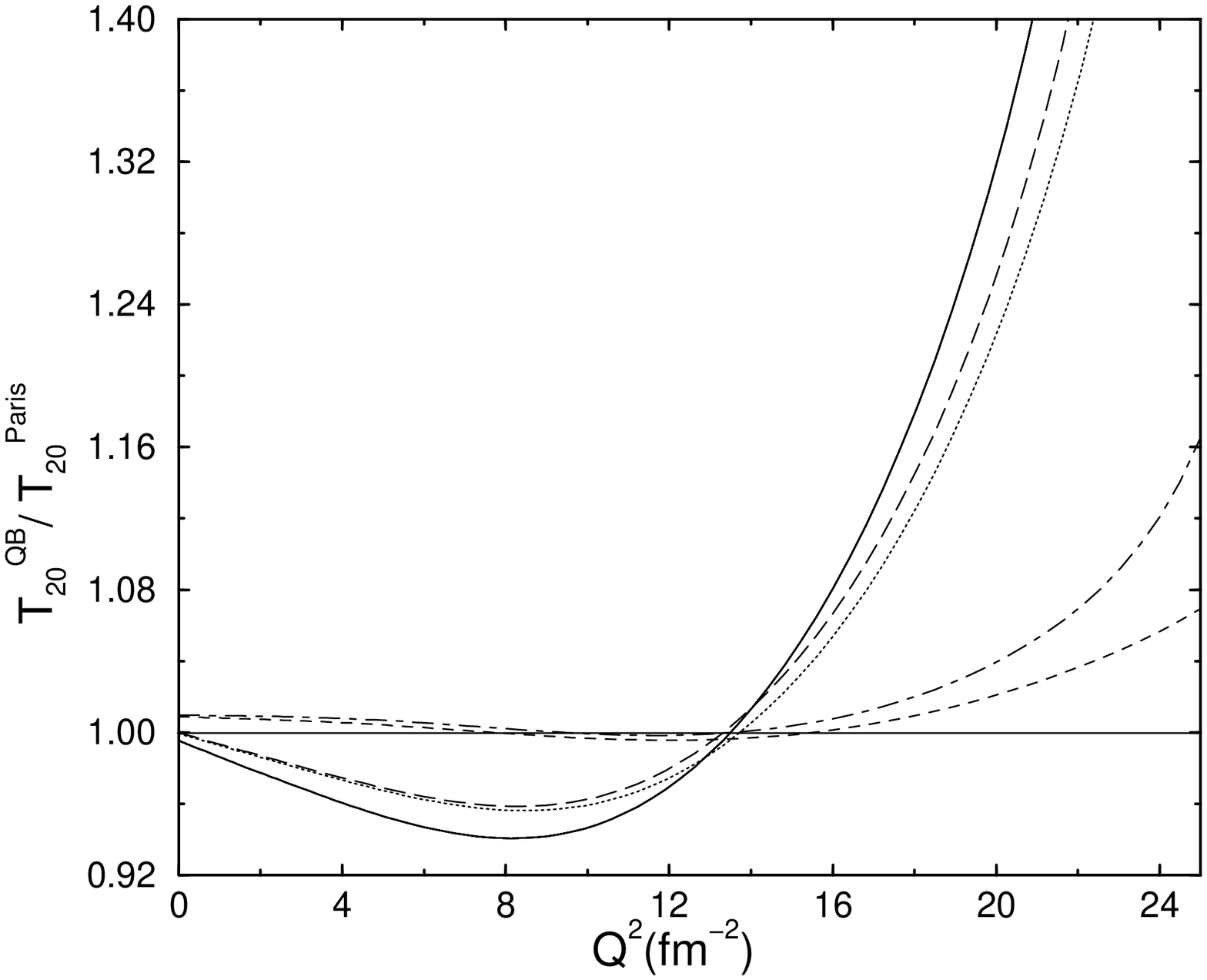, angle=270, width=5.5cm} }  
\caption{Ratio of predictions for the Bonn-QB and Paris models: effect of 
non-local terms in the interaction (see text for explanations)\label{fig1}}
\end{center}
\end{figure} 

Ratios of form factors and related quantities calculated in different 
representations of the interaction should go to 1 when the 
non-locality effect is accounted for and provided that the unitary 
equivalence is ensured. 
It is not so in practice because the unitary transformation is treated 
at first order. In each case, there are thus two sets of results, 
depending on whether corrections are added to predictions made with one 
model or removed from the other ones.

\begin{figure}[htb]
\begin{center}
\mbox{  \epsfig{ file=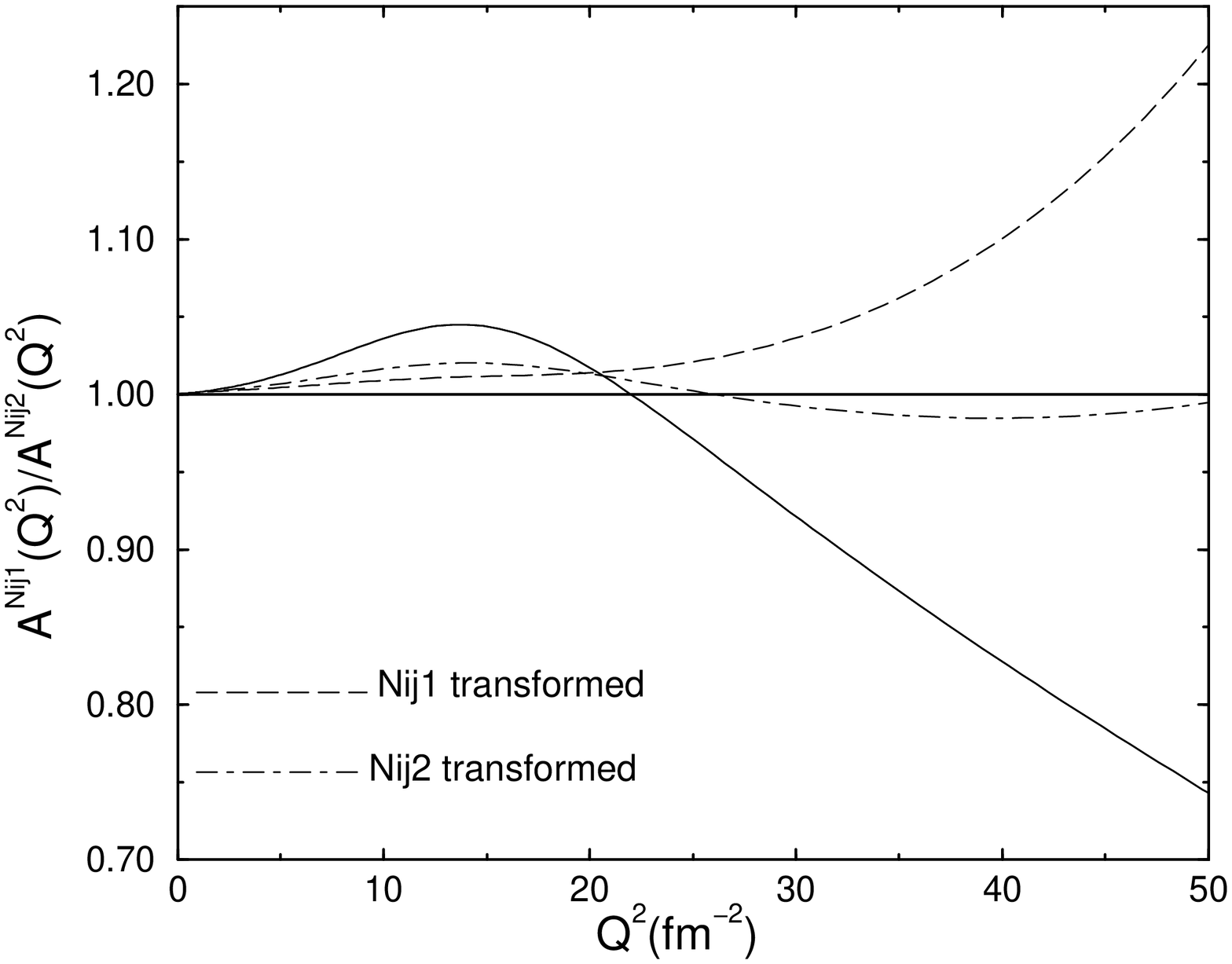, angle=270, width=5.5cm}
\hspace*{0.5cm} \epsfig{ file=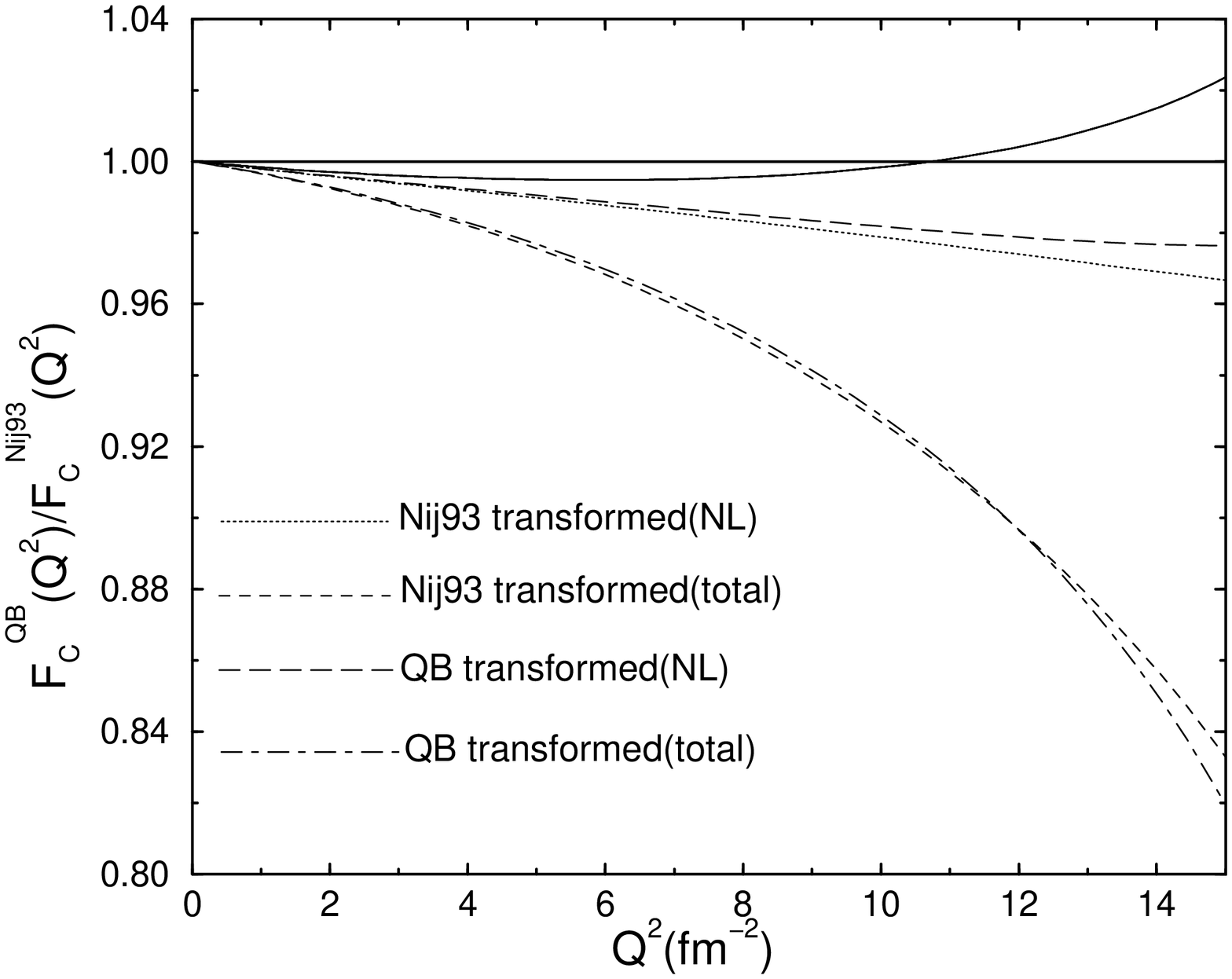, angle=270, width=5.5cm} }  
\caption{Examples showing a decrease of discrepancies between models due a 
scalar non-local effect (left part) and the appearance, on the contrary, of a 
discrepancy (right part); see text for comments\label{fig2}}
\end{center}
\end{figure}

\begin{figure}[htb]
\begin{center}
\mbox{ \epsfig{ file=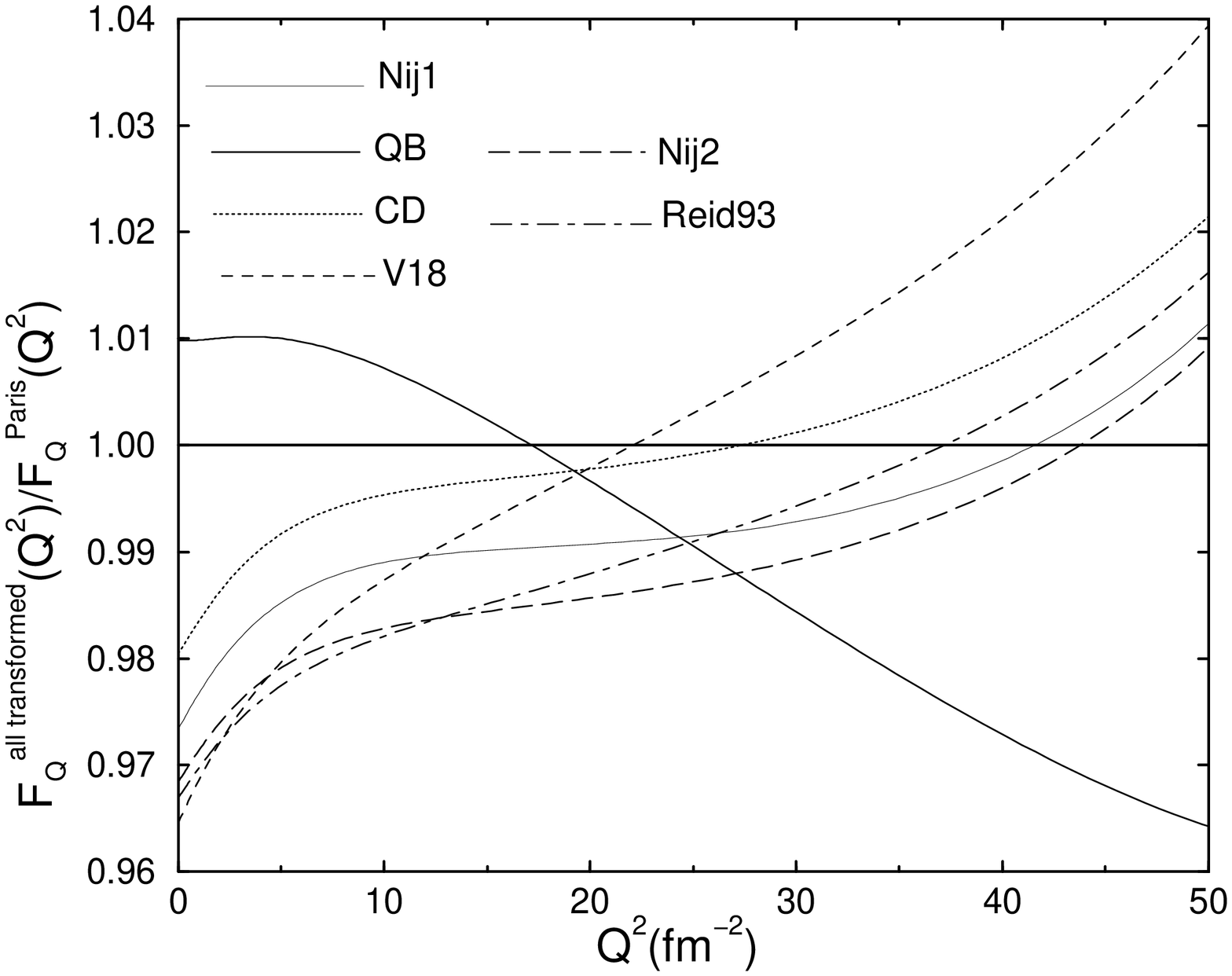, angle=270, width=5.5cm} 
\hspace{0.5cm}  \epsfig{ file=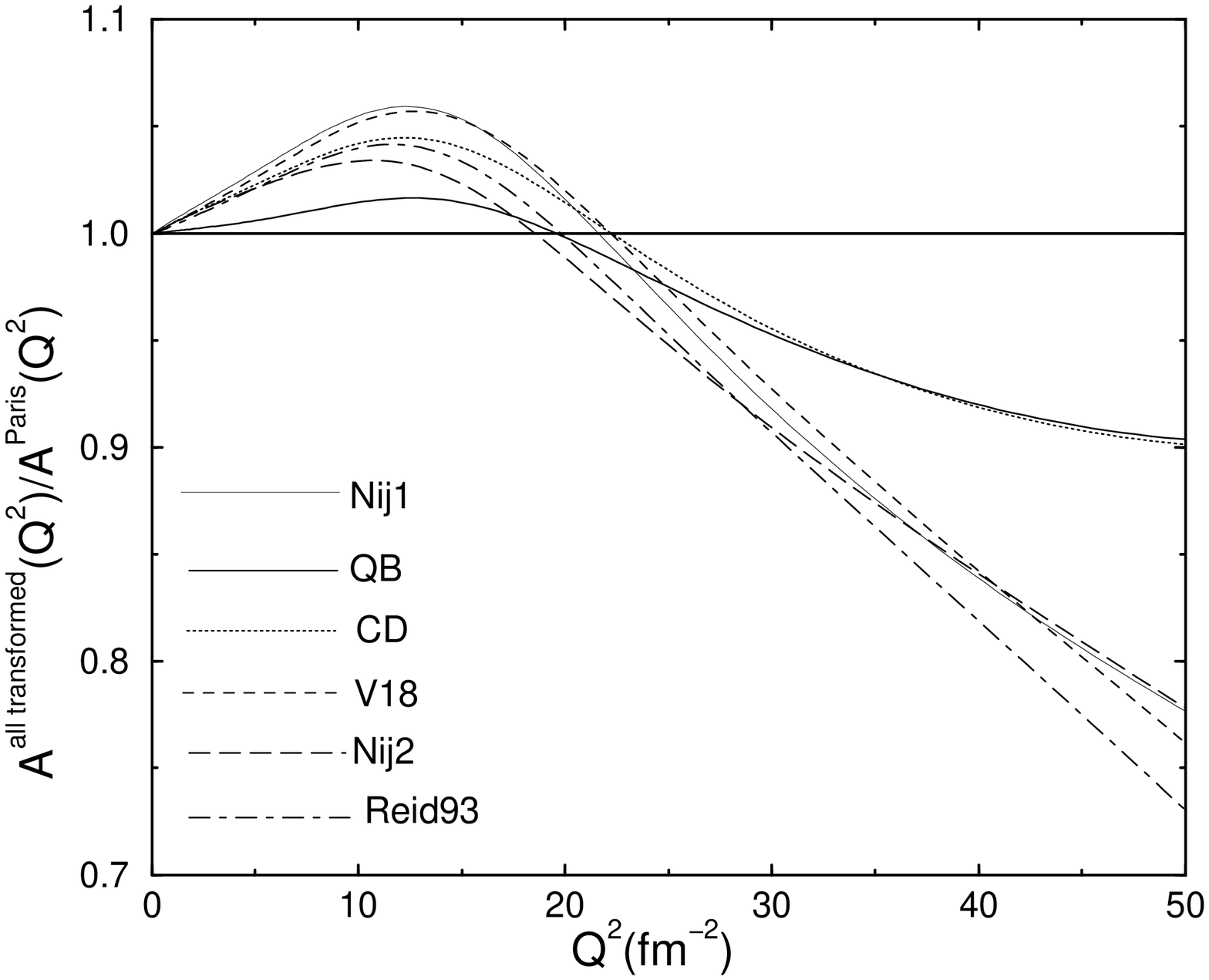, angle=270, width=5.5cm} }  
\caption{Comparison of predictions of different models after incorporating 
non-locality corrections (see comments in the text)\label{fig3}}
\end{center}
\end{figure}

Ratios of predictions made from the Bonn-QB and Paris models for the $A(Q^2)$  
and $T_{20}(Q^2)$ observables are shown in Fig. \ref{fig1}. As it can be seen, 
the ratio of bare predictions (continuous line) tends to 1 when the effect of 
the anticommutator (dashed and dotted lines) and commutator terms (small-dash 
and dashed-dotted lines) is considered. A large part of the effect is due to the 
tensor part of $V_{NL}$. Notice that a slight departure from 1 appears for 
$T_{20}(Q^2)$ around $Q^2=0$.
The effect of the scalar part of the anticommutator, which is seen in Fig. 
\ref{fig2} (left part), shows features similar to the previous ones. It involves 
the S-wave function close to the origin, which is generally suppressed in local 
models compared to non-local ones. The right part of the same figure emphasizes 
a case where an agreement between two models turns into a disagreement. In fact, 
this one is consistent with what is expected from the comparison of the 
asymptotic normalization, unaffected by the present analysis. In Fig. 
\ref{fig3}, all predictions corrected for the effect of a non-local term 
are compared to the Paris ones. At low $Q^2$, it is seen that some 
discrepancy is still present while the initial motivation of the work was rather 
to explain it by non-locality effects. Actually, for both  $A(Q^2)$  and 
$T_{20}(Q^2)$, the discrepancy reflects a sensitivity to the $A_D/A_S$ ratio in 
the first case and $A_S$ in the second one. Around $Q^2=20\; {\rm fm}^{-2}$, the 
ratio becomes very close to 1, while in absence of corrections departures up to 
10\% and 20\% respectively could be observed. The decrease of the uncertainty 
has motivated Schiavella and Sick in using the quadrupole form factor to derive 
the neutron charge form factor\cite{SICK}.

\section{Conclusion}
Effects of non-local terms in NN interaction models have been considered. 
Roughly, they explain a large part of the differences that the comparison of 
various model predictions for electromagnetic observables evidences. The 
part with a tensor character has been found to be the dominant one. It is 
also the best determined. Similar conclusions hold in some cases for the scalar 
part but the effect is often masked by other effects related to the fact that 
models correspond to different values of the asymptotic normalization, $A_S$, 
which is an observable and remains unchanged in the analysis performed here. 

While the original goal of present studies was rather to relate discrepancies 
between models to some non-locality, it appears that this is not always 
possible, especially at low $Q^2$. Interestingly however, accounting for this 
effect tends to restore some hierarchy of the results, as expected from simple 
models. Thus, in the above range, the structure function, $A(Q^2)$, and the  
quadrupole form factor, $F_Q(Q^2)$, evidence a direct sensitivity respectively 
to the asymptotic normalizations, $A_S$ and $A_D$, which are unaffected by the 
present analysis of non-local effects. 

The result of the analysis presented here was not a priori guaranteed. The fact 
that it points to a unique family of phase-equivalent models indicates that the 
sensitivty of the models to different parametrizations of the radial part for 
instance or to a different fit to experimental data is rather small. Thus, the 
availability of various models is not without interest. The remaining 
sensitivity, as for $A_S$ or equivalently the scattering length, $a_t$, 
strongly calls for a more accurate determination of these quantities. 

Throughout the present study, we compare together predictions of models for 
electromagnetic observables. Ultimately, 
a comparison to experiment should be done. In this respect, a model to be 
prefered, most probably non-local, is that one based on degrees of freedom 
of which effectice character, unavoidable in any case, is as low as possible.

\end{document}